\documentclass[a4paper,fleqn,usenatbib]{mnras}
\usepackage{amssymb,amsmath,graphicx,psfrag,mathtools}
\usepackage[T1]{fontenc} 
\usepackage{ae,aecompl} 
\usepackage{url}
\title[Coherent Plasma-Curvature Radiation in FRB]{Coherent Plasma-Curvature Radiation in FRB}
\author[J. I. Katz]{
J. I. Katz,$^{1}$\thanks{E-mail katz@wuphys.wustl.edu} 
\\
$^{1}$Department of Physics and McDonnell Center for the Space Sciences,
Washington University, St. Louis, Mo. 63130 USA 
}
\date{Accepted XXX.  Received YYY; in original form ZZZ} 
\pubyear{2017} 
\date{\today}
\begin{document} 
\label{firstpage} 
\pagerange{\pageref{firstpage}--\pageref{lastpage}} 
\maketitle 
\begin{abstract}
Curvature radiation is a natural candidate for the emission mechanism of
FRB.  However, FRB spectra have structure with $\Delta \nu/\nu \sim 
\text{0.03--0.2}$, inconsistent with the very smooth spectrum of curvature
radiation.  Although this spectral structure might be attributed to
chromatic scintillation {or lensing}, in four FRB high spectral
resolution data indicate scintillation decorrelation bandwidths much
narrower than the observed $\sim \text{30--300}$ MHz spectral structure.
{Some of} the observed structure {may} be intrinsic to the radiation
mechanism.  I suggest that the observed spectral structure reflects the
spatial structure of a clumpy radiating charge distribution, and that the
characteristic curvature radiation frequency {may be} higher than the
observed frequencies.  In this coupled plasma-curvature radiation process
the radiated spectra are the product of the spectra of the plasma wave and
that of incoherent curvature radiation.  The argument applies to all
coherent radiation processes, including those that produce pulsar nanoshots.
The implied FRB ``clump'' charges are large, and produce electrostatic
potentials that suggest electron Lorentz factors $\gtrsim 10^2$.  The result
applies generally to coherently radiating sources.
\end{abstract}
\section{Introduction}
The short durations and high radiated energy of Fast Radio Bursts (FRB)
point to neutron stars, with their small size and deep gravitational
potential wells, as their origin.  The high brightness of FRB requires
coherent radiation by ``clumps'' with net charges of multiple Coulombs
\citep{K14,K16}.  A magnetic field { provides} possible energy sources,
{ including reconnection of magnetostatic fields and magnetic dipole
radiation and particle acceleration powered, as in radio pulsars, by
rotational energy.  It enables the emission of radio-frequency curvature
radiation.  Plasma instability powered by counterstreaming electrons and
positrons provides additional radiation mechanisms.}

Coherent curvature radiation has been often suggested as the radiation
mechanism of FRB \citep{DWWH16,GDLMW16,WYWDW16,GL17,K17b,YZ17} because in
the intense fields of neutron star magnetospheres any energy of gyration
about field lines is very quickly radiated as X-rays or gamma-rays, leaving
the particles to follow the field lines.  \citet{Me17} presents a general
review of coherent radiation processes and \citet{KLB17,LK17} present a
very detailed investigation in the context of FRB.

The spectra of FRB at frequencies in the 1.2--1.5 GHz band and, in a few
observations, at frequencies from 700 MHz to 8 GHz, contain structure on
frequency scales $\sim$ 30--300 MHz \citep{Ke12,T13,BSB14,S14,M15,P15,R15,R16,Sc16,Sp16,B17,C17,H17,L17,M17,O17,P17,G18,M18}.
Spectral structure could not be seen in the discovery paper \citep{L07}
because the dynamic spectral data were { saturated.}

{\citet{CWH17} attribute the complex temporal (on $\sim 100\,\mu$s time
scales) and spectral structure of FRB \citep{R16,G18,M18} to chromatic
lensing, but the observed structure does not closely resemble the
predictions of their Figs.~6 and 7.  In this paper I consider an alternative
explanation, that these properties are intrinsic to the plasma emission
mechanism.}

This behavior is widespread.  Pulsar nanoshots have analogous frequency
structure with $\Delta \nu \sim 0.1 \nu$ \citep{S04,HE07}.  These nanoshots
are, scaled up by many orders of magnitude, a popular model for FRB. 

These observed spectra are inconsistent with the spectra emitted by point
charges traveling on curved paths at constant speed (acceleration 
perpendicular to velocity), and hence with a straightforward interpretation
as curvature or synchrotron radiation \citep{K14}.  Such particles radiate
very smooth spectra \citep{J99}, with a gradual rolloff between a low
frequency asymptote $F(\omega) \propto \omega^{1/3}$ that is a weak function
of frequency and a high frequency asymptote $F(\omega) \propto \omega^{1/2}
\exp{(-\omega/\omega_c)}$ that falls off rapidly with increasing frequency.
The characteristic frequency $\omega_c = 3 c \gamma^3/(2 \rho)$, where
$\gamma$ is the particle's Lorentz factor (taken as constant) and $\rho$ the
radius of curvature of its path.  The same expression applies to synchrotron
radiation as to curvature radiation, except that for curvature radiation
$\rho$ is given by the magnetic field geometry while for synchrotron
radiation it is given by the radius of the charge's circular or helical
path.

\citet{K14} explained these observations by attributing the radiation to
emission by plasma waves, modeling them as electric dipole radiators.  In
this paper I suggest a hybrid plasma-curvature radiation process that
recognizes that FRB are likely produced in regions of intense and curved
magnetic fields.  This process resolves the disagreement between the
observed structured FRB spectra and the theoretical spectrum of curvature
radiation.  I first consider, and argue against, the hypothesis that the
frequency structure is the result of chromatic scintillation along the
propagation path.  I then discuss the effect of the spatial structure of
coherently radiating charge densities on the emitted spectrum, and argue
that plasma-curvature radiation may explain the observed properties of FRB.
{ \citet{GL17} discuss the broad, rather than high resolution, frequency
structure of coherent curvature radiation, involving considerations other
than the plasma waves suggested here.}
\section{Scintillation}
Can the frequency structure of FRB spectra be attributed to scintillation?
Strong scintillation produces a Rayleigh distribution of received energy in
non-overlapping temporal or spectral intervals.  Data from temporal
intervals are less useful because scintillation is less readily
distinguished from variation in the source emission power (although the
chromaticity of plasma scintillation might enable this distinction and there
is no reason to expect intrinsic variations in emitted power to follow a
Rayleigh distribution) and because bursts may be resolved into only a few
temporal bins.  In addition, temporal resolutions have been $\gtrsim 0.03$
ms, making it impossible to set upper limits on scintillation bandwidths
$\gtrsim 30$ kHz.
\subsection{Narrow Band Spectral Structure}
Spectrally resolved data are more useful.  The scintillation decorrelation
bandwidths of four FRB have been measured or bounded:

\citet{M15} found a spectral decorrelation width of $1.2 \pm 0.4$ MHz in
FRB 110523 that { also} has spectral structure on a scale of $\sim$ {
10--30 MHz (their Fig.~1)} around
a frequency of 800 MHz ({\it cf.\/} the Stokes $I$ data in their Fig.~2 and
the spectral autocorrelation function in their Extended Data Fig.~3).

\citet{R16} found a decorrelation width (possibly interpretable as an upper
limit) of $100 \pm 50$ kHz for FRB 150807.  Fig.~S9 of this paper shows a
quantitative fit to a Raleigh distribution, from which { saturated
scintillation may be inferred} and the decorrelation bandwidth derived.
This cannot { itself} explain the spectral structure evident on frequency
scales of $\sim \text{30--100}$ MHz.  { The spectro-temporal data in
their Fig.~1 indicate that the frequency and temporal dependence on this
frequency scale are not separable; with scintillation essentially constant
through this sub-ms FRB,} the spectral structure must be attributed to the
emission mechanism.

\begin{figure}
	\centering
	\includegraphics[width=0.99\columnwidth]{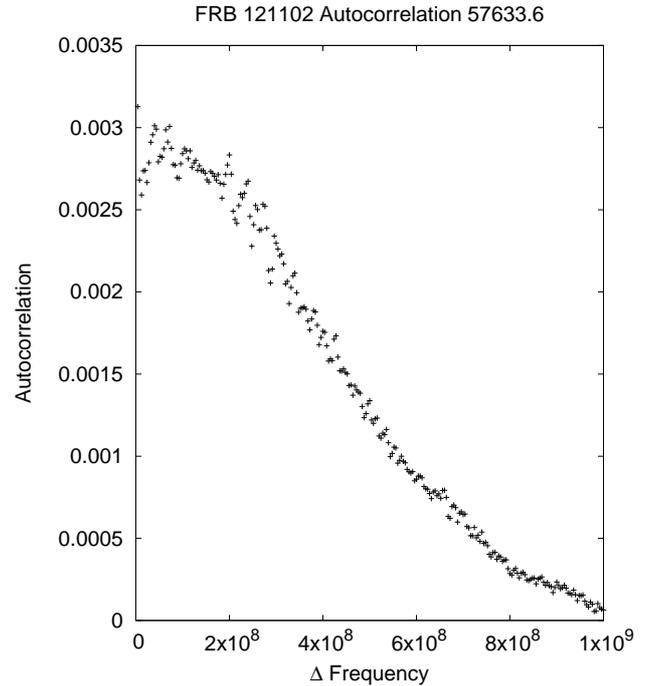}
	\caption{\label{57633correl}Spectral autocorrelation function of
	pulse 57633.6 from FRB 121102, observed at the VLA \citep{L17}.
	Normalization is arbitrary.
	The gradual rolloff reflects the $\sim 250$ MHz spectral width,
	while the absence of significant structure near the resolution of
	4 MHz indicates a decorrelation width $\lesssim 4$ MHz.}
\end{figure}

The VLA observations \citep{L17} of pulse 57633.6 (MJD 57633.67986367) of
FRB 121102 around 3 GHz (their Fig.~4) suggest a Rayleigh distribution of
energy, although a quantitative test is not possible because the underlying
smooth spectral distribution is not known; their Gaussian is only a fit.
Such a Rayleigh distribution would imply a decorrelation bandwidth less than
the spectral resolution (4 MHz) of the data.  The autocorrelation function
is shown in Fig.~\ref{57633correl} and indicates an upper bound on the
decorrelation bandwidth of $\lesssim 4$ MHz.  The expected Galactic
scintillation bandwidth is $\sim 7$ MHz \citep{CL02,L17}; if the full
(including near-source contributions) scintillation bandwidth is $< 4$
MHz it explains the observed decorrelation. 

\citet{F18} found a spectra decorrelation width of about 1.5 MHz in FRB
170827 with substructure on scales of 100--200 kHz in UTMOST observations
over a bandwidth of 31.25 MHz centered at 835 MHz.  Although the observing
bandwidth is narrow, compared to the Parkes bandwidth of 288 MHz centered at
1372.5 MHz \citep{L07}, there appears to be structure on frequency scales
broader than the 1.5 MHz decorrelation width.

\subsection{Broad Band Spectral Structure}
{ There is also spectral structure on broader scales of tens to hundreds
of MHz.  This is shown in ``waterfall'' plots for FRB 110220 \citep{T13} and
many other FRB \citep{FRBCat}, in Fig.~1 of \citet{R16} for FRB 150807, and
and for FRB 121102 by \citet{L17,M18,G18}.  \citet{Sp16} have argued that the
varying spectral shapes of bursts from FRB 121102 are intrinsic, an argument
I develop futher here.

The broader frequency structure of FRB 121102 is difficult to explain as the
result of scintillation.  It varies greatly from burst to burst (Fig.~2 of
\citet{G18}) and even on sub-ms time scales within bursts (\citet{G18} and
Figs.~1 and ED1 of \citet{M18}).  It is implausible that a scattering
screen could change that rapidly (for Galactic pulsars the scintillation
decorrelation time is minutes \citep{R90}), and in the nanoshots of the Crab
pulsar the spectrum changes on $\sim \mu$s time scales \citep{HE07}, an even
more demanding condition.  If it were attributed to a slowly (compared to
$\sim 1$ ms FRB durations) varying frequency-dependent scintillation pattern
$f(\nu)$ then the complete spectro-temporal intensity function would be
separable: $F(\nu,t) = f(\nu)g(t)$ with a temporal modulation $g(t)$, which
is clearly not the case for FRB 121102 \citep{L17,M18,G18} although it may
be valid for FRB 170827 \citep{F18}.  In addition, the distribution of
intensities of this broad-band structure is unlike the Rayleigh statistics
of strong scintillation that are observed (Fig.~S9 of \cite{R16}) at
frequency resolutions of a few MHz or less; Fig.~4 of \citet{L17} shows an
excess of broad-band averaged intensities consistent, at high $S/N$, with
zero that would be rare for Rayleigh statistics.

This broad frequency structure plausibly explained as rapid variation in the
radiated spectrum.  Such complex and rapidly changing spectra are very
different from the smooth spectrum of accelerated point charges, either
elementary charges or charges correlated in charge ``bunches'' that radiate
as macro-point charges, and require explanation.}
\section{Radiation by a Continuous Distribution of Charge Density}
\subsection{Spectrum}
The radiation field produced by a charge moving at constant speed $\beta c$
on a path locally approximated by a circular arc is a familiar classical
result \citep{J99}.  Let $\theta(t_r) = \beta c t_r/\rho$ denote the
position of the charge on the arc at the retarded time $t_r$, $\beta c$ is
the particle's speed and $\rho$ is the radius of curvature of the arc.  The
observer is at a distance $d$ ($d \to \infty$) in a direction that makes an
angle $\alpha$ to the plane of the osculating circular arc.  Then the
component of electric field perpendicular to the plane of the arc produced
by a point charge $q$ is \citep{J99,KLB17}
\begin{equation}
\label{Eq}
E_{point}(t_{obs}) = {q \beta^2 \over d \rho}
\left[{\cos{\theta(t_r)} - \beta\cos{\alpha} \over
(1 - \beta\cos{\theta(t_r)}\cos{\alpha})^3}\right],
\end{equation}
where $t_r$ is related to the observer's time $t_{obs}$ by
\begin{equation}
t_{obs} = t_r - {\rho \cos{\alpha} \over c} \sin{\beta c t_r \over \rho}.
\end{equation}
This transcendental equation for $t_r(t_{obs})$ is readily solved
numerically.  $E_{point}(t_{obs})$ has the Fourier transform
$E_{point,\omega}$, the familiar amplitude spectrum of radiation by an
accelerated particle.

As calculated explicitly by \citet{KLB17}, this component of electric field
has a non-zero mean (although it does change sign as the charge moves on its
path), while the component in the plane of the orbit averages to zero.  Both
components have widths $\sim \rho/\gamma^3 c$, where $\gamma \gg 1$ is
the particle's Lorentz factor, corresponding to the characteristic frequency
$\omega_c \sim \gamma^3 c/\rho$.  Both components of field contribute to the
radiation (that is strongly linearly polarized in this simple geometry), but
it is sufficient to consider only one component.

Emission as bright as that of FRB requires enormous net charges radiating
coherently \citep{K14} and the incoherent emission of individual elementary
charges (likely electrons and positrons) may be neglected.  If there {
is} a continuous distribution of net charge along the orbit
$\lambda(\theta)$ each element of charge $dq = \lambda(\theta)d\theta$
contributes independently.  The integral $\int\!dq$ must be small or zero
(because otherwise a large electrostatic field would produce neutralizing
currents) with near cancellation of regions of positive and negative charge;
large local charge densities $\vert \lambda(\theta) \vert$ produce the
intense radiation. 

We assume that the charge density $\lambda(\theta)$ moves uniformly along
the circular arc; $\lambda$ is static in a frame moving at the speed $\beta
c$ and angular velocity $\beta c/\rho$.  This assumption is not necessary,
but leads to a simple result.  The total electric field is found by
integrating the contributions of each $dq$:
\begin{equation}
E_{tot}(t_{obs}) \propto \int\!E_{point}(t_{obs}-\Delta t)
\lambda(\Delta \theta)\,d\Delta \theta
\end{equation}
where $\Delta \theta = \beta c \Delta t/\rho$.

The spectrum of radiated power $P(\omega)$, is observed.  By the Convolution
Theorem
\begin{equation}
\label{power}
P(\omega) \propto \left| E_{tot,\omega} \right|^2 \propto
\left|E_{point,\omega}\right|^2 \left|\lambda_\omega\right|^2.
\end{equation}
$\vert E_{point,\omega} \vert^2$ is proportional to the familiar spectrum
$F(\omega)$ of curvature or synchrotron radiation, increasing $\propto
\omega^{1/3}$ for $\omega \ll c \gamma^3/\rho$ and falling nearly
exponentially for $\omega > c \gamma^3 / \rho$.  $\lambda_\omega$ is the
Fourier transform of $\lambda(\theta)$ with $\theta = \beta c \Delta t/
\rho$.  {Eq.~\ref{power} relates the observed spectrum $P(\omega)$ to
the charge density distribution $\lambda_\omega$.  The spatial scale of
charge structure $\rho\theta \lesssim \gamma^3 \beta c/\omega$.  The
observation of spectral structure on a scale $\Delta \omega \ll \omega$
implies comparatively narrow-band spatial structure with a wave-number
width $\Delta k \sim (\Delta \omega /\omega) k \sim \Delta \omega/(\gamma^3
\beta c)$.}

For a point charge $\lambda(\theta) = q \delta(\theta-\theta_0)$,
$\lambda_\omega = \text{Constant}$ and $P(\omega)$ is the usual spectrum
of radiation by a single charge.  For a uniform distribution of charge
$\lambda(\theta) = \text{Constant}$, $\lambda_\omega \propto \delta(\omega)$
and there is only a stationary field, without radiation.

The charge density distribution, with power spectrum $\vert \lambda_\omega
\vert^2$, is a result of the plasma process that must produce the
``clumps'', and is unknown.  It is likely to be cut off
above some frequency $\omega_p$, perhaps an electron plasma frequency.  A
process that accelerates electrons also accelerates positrons in the opposite
direction, naturally leading to two-stream instability with frequencies
close to the electron plasma frequency \citep{LK17}.

If $\omega_p \gg \omega_c$ then the radiated spectrum $P(\omega)$ will be
cut off exponentially by $E_{point,\omega}$ for $\omega > \omega_c$ and will
be modulated by $\vert \lambda(\omega) \vert^2$ for $\omega < \omega_c$.  If
$\omega_p \ll \omega_c$ the radiation spectrum will be $\propto \omega^{1/3}
\vert \lambda_\omega \vert^2$.  This is essentially the spectrum of the
charge density passing the point on the arc ($\theta = 0$) at which the
motion is most closely directed towards the observer.  The emitted radiation
has the spectrum of the plasma turbulence {multiplied by} the intrinsic
spectrum $E_{q,\omega}$ of curvature radiation by a point particle.
{Because the latter spectrum is so broad and smooth, at frequencies less
than the characteristic curvature radiation frequency their product is
close to the spectrum of the turbulence}.  It is then not possible to infer
$\gamma$ from the observed frequency $\omega_{obs}$ except to set a lower
bound from the condition that $\omega_{obs} \lesssim \omega_p \ll \omega_c$:
\begin{equation}
\label{gammacurve}
\gamma \gtrsim \left({\omega_{obs} \rho \over c}\right)^{1/3} \approx 50
\end{equation}
for neutron star parameters and the observation \citep{G18} of FRB 121102 at
$\nu = 8$ GHz. For FRB observed at 1.2--1.5 GHz the corresponding lower
bound is $\gamma \gtrsim 30$.  Without an understanding of the plasma
turbulence, it is not possible to predict to how high frequencies FRB
produce observable coherent radiation.
\subsection{The Clump Charge}
It is possible to estimate the coherently moving charges $Q$ required to 
explain the observed FRB.  Using Eq.~\ref{Eq} \citep{KLB17}, the maximum
electric field at the observer is obtained by taking $\alpha = 0$ and
$\theta = 0$:
\begin{equation}
\label{Emax}
E_{max} = {Q \beta^2 \over d \rho} \left[1 - \beta \over
(1 - \beta)^3\right] \approx {Q \over d \rho}{1 \over (1 - \beta)^2}
\approx {4 Q \gamma^4 \over d \rho},
\end{equation}
where we take $\gamma \gg 1$.  Fields comparable to this value are radiated
into a solid angle $\sim \gamma^{-2}$ at the source.

An isotropically radiating source of radius $R$ will have $\sim R^2 /
[\gamma^2 (\lambda/2)^2]$ independently radiating emitters producing fields
$\sim E_{max}$ at the observer, provided $\gamma < 2R/\lambda$.  These
emitters combine incoherently, and their intensities add.  Then, using
\begin{equation}
I = {R^2 \over \gamma^2 (\lambda/2)^2} {c \over 8 \pi} E_{max}^2
\end{equation}
and taking $R \sim \rho$, as expected for a dipole field,
\begin{equation}
\label{Q}
Q \sim \sqrt{\pi I \over 8 c}{\lambda d \over \gamma^3} \sim \sqrt{I \over
\text{Jy-GHz}}{\lambda \over \text{20 cm}}{d \over \text{1 Gpc}}{1 \over 
\gamma^3}\ 2 \times 10^{16}\ \text{esu}.
\end{equation}
If $\gamma \sim 60$, as required for $\omega_c = 2 \pi \times 1.4$ GHz
{curvature} radiation {from a neutron star magnetosphere}, and $I
\sim 1$ Jy-GHz, a nominal FRB value, then $Q \sim
10^{11}\ \text{esu} \sim 30\ \text{Coulombs}$.  This value of $\gamma$ is
only a rough lower bound (\citet{KLB17} consider $\gamma \sim 30$).  If
$\gamma$ is higher the required $Q$ is less; then Eq.~\ref{Emax} must be
replaced by a more complex expression \citep{J99}.  Although the observation
of radiation at some frequency $\omega$ places a lower bound on $\gamma$,
it may exceed this lower bound by a large factor (if $\omega \ll \omega_c$),
so that Eq.~\ref{Q} cannot, without independent information about $\gamma$,
be interpreted as placing a lower bound on $Q$.

These results also apply to observers illuminated by a collimated source
\citep{K17a,K17c}.  Collimation reduces the total radiated power, but does
not change the required $Q$.
\subsection{Characteristic Lorentz Factor and Frequency}
{ Taking} $\gamma^3 = 2 \omega_c \rho /(3 c) = (4 \pi/3) \rho/\lambda$
and $\rho \sim 10^6$ cm { and substituting in Eq.~\ref{Q}}:
\begin{equation}
Q \sim \sqrt{I \over \text{Jy-GHz}} \left({\lambda \over \text{20 cm}}
\right)^2 {d \over \text{1 Gpc}}\ 4 \times 10^{10}\ \text{esu},
\end{equation}
Qualitatively, higher frequency radiation may make less severe demands on
$Q$.

The existence of large coherent charge clumps (Eq.~\ref{Q}) implies, {if
the charge is distributed over a region of size $\sim \lambda/2$,} large
electrostatic potentials
\begin{equation}
\label{V}
V \sim {Q \over (\lambda/2)} \sim \sqrt{\pi I \over 2 c} {d \over \gamma^3}.
\end{equation}
This suggests a minimum or characteristic Lorentz factor $\gamma_c$ set by
equating the electron or positron energy $\gamma_c m_e c^2$ to its
electrostatic energy $eV$.  Then
\begin{equation}
\label{gammac1}
\begin{split}
\gamma_c &= \left({\pi I \over 2c}\right)^{1/8}
\left({ed \over m_e c^2}\right)^{1/4}\\ &\sim
\left({I \over \text{Jy-GHz}}\right)^{1/8}
\left({d \over \text{Gpc}}\right)^{1/4}\ 1.1 \times 10^3.
\end{split}
\end{equation}
Defining an equivalent isotropic radiated power $P_{eq} \equiv 4 \pi I d^2$,
this may be rewritten
\begin{equation}
\label{gammac2}
\gamma_c = \left({P_{eq} \over 8 m_e^2 c^5/e^2}\right)^{1/8} \approx
1000\ \left({P_{eq} \over 10^{42}\ \text{erg/s}}\right)^{1/8}.
\end{equation}

{The fields of collimated relativistic particles moving towards the
observer add coherently if they are distributed over a distance along their 
path $\sim \gamma^2 \lambda/2$.  Then Eqs.~\ref{V}--\ref{gammac2} are
replaced by
\begin{equation}
\label{V2}
V \sim {Q \over (\lambda/2) \gamma^2} \sim \sqrt{\pi I \over 2 c}
{d \over \gamma^5}.
\end{equation}
Again equating $\gamma_c m_e c^2 = eV$,
\begin{equation}
\label{gammac12}
\begin{split}
\gamma_c &= \left({\pi I \over 2c}\right)^{1/12}
\left({ed \over m_e c^2}\right)^{1/6}\\ &\sim
\left({I \over \text{Jy-GHz}}\right)^{1/12}
\left({d \over \text{Gpc}}\right)^{1/6}\ 1.0 \times 10^2
\end{split}
\end{equation}
and
\begin{equation}
\label{gammac22}
\gamma_c = \left({P_{eq} \over 8 \pi m_e^2 c^5/e^2}\right)^{1/12} \approx
100\ \left({P_{eq} \over 10^{42}\ \text{erg/s}}\right)^{1/12}.
\end{equation}
These estimates of $\gamma_c$ are more general than that of
Eq.~\ref{gammacurve} that refers specifically to curvature radiation in
inner neutron star magnetospheres.}

These results apply to any radiation process in which coherent charge
bunches are accelerated perpendicular to their velocity, and even if the
radiation is collimated and the actual radiated power is much less than
$P_{eq}$.  Note that it is possible that the Lorentz factor of the radiating
charges $\gamma \gg \gamma_c$; $\gamma_c$ is only the minimum implied by the
existence of large electrostatic potentials in the source.

The characteristic Lorentz factor of Eq.~\ref{gammac1} implies a
characteristic frequency of curvature radiation $\omega_c = 3 \gamma_c^3
c/2\rho$:
\begin{equation}
\label{omegac}
\omega_c \sim \left({I \over \text{Jy-GHz}}\right)^{3/8}
\left({d \over \text{Gpc}}\right)^{3/4}
\left({10^6\ \text{cm} \over \rho}\right)\ 
5 \times 10^{13}\,\text{s}^{-1},
\end{equation}
{while Eq.~\ref{gammac12} implies
\begin{equation}
\label{omegac2}
\omega_c \sim \left({I \over \text{Jy-GHz}}\right)^{1/4}
\left({d \over \text{Gpc}}\right)^{1/2}
\left({10^6\ \text{cm} \over \rho}\right)\ 
5 \times 10^{10}\,\text{s}^{-1},
\end{equation}
}
If the factors in parentheses are $\sim 1$, {as for a FRB produced in a
neutron star magnetosphere at cosmological distances}, Eq.~\ref{omegac}
corresponds to infrared radiation with $\lambda \sim 30\,\mu$ {while
Eq.~\ref{omegac2} corresponds to $\nu \sim 10$ GHz, approximately the
highest frequency at which FRB have been observed}.  For there to be coherent
radiation at these frequencies, the structure factor $\lambda_\omega$ of the
charge distribution must be significant for $\omega \sim \omega_c$.  The
underlying plasma physics is complex and likely not understood, and there is
no {\it a priori\/} reason to expect observable coherent radiation.  Despite
this, because $\gamma_c$ is only a lower bound curvature radiation at yet
higher frequencies than those of Eqs.~\ref{omegac} and \ref{omegac2} is
possible.
\section{Discussion}
The classical spectrum of curvature or synchrotron radiation \citep{J99} is
that radiated by point charges.  Coherent emission requires a large number
of charges spread over a finite region; electrostatic repulsion spreads out
``clumps'' of charge that cannot be treated as points.  They must be
described by continuous distributions of charge density, which also imply
continuous distributions of the times they pass the small (for relativistic
particles) region producing significant fields at the observer.

Only the insignificant incoherent part of the emission is described by the
classical result $E_{point,\omega}$ for the radiated field of point charges.
Very little power may be emitted at frequencies around the classical
characteristic frequency $\omega_c$ of incoherent curvature radiation.  The
Lorentz factors of radiating particles cannot then be inferred from the
frequency of the observed radiation and may be much higher.  Observed
spectral cutoffs are likely to reflect the spatial distribution of the
charge density, not the characteristic frequency of curvature radiation.

The coherently radiated field reflects the distribution and motion of the
smoothed charge density, a continuous function of the coordinates.  The
radiation mechanism may be described as hybrid coherent plasma-curvature
radiation.  This resolves the disagreement between the physically plausible
mechanism of curvature radiation and the apparently inconsistent observed
spectrum.

The results of Eqs.~\ref{gammac1}--\ref{omegac} are very generally
applicable to coherent nonthermal sources of high brightness emission, not
only FRB, and do not depend on the emission mechanism.
\section*{Acknowledgements}
I thank C. J. Law for providing the raw data for Fig.~\ref{57633correl} and
V. Gajjar, P. Kumar, C. J. Law, W. Lu and V. Ravi for useful discussions.

\bsp 
\label{lastpage} 

\begin{thebibliography}{99}
\bibitem[\protect\citeauthoryear{Bannister {\it et al.\/}}{2017}]{B17}
Bannister, K. W., Shannon, R. M., Macquart, J.-P. {\it et al.\/} 2017 \apjl\
841, L12.
\bibitem[\protect\citeauthoryear{Burke-Spolaor \& Bannister}{2014}]{BSB14}
Burke-Spolaor, S. \& Bannister, K. W. 2014 \apj\ 792, 19.
\bibitem[\protect\citeauthoryear{Chatterjee {\it et al.}}{2017}]{C17}
Chatterjee, S., Law, C. J., Wharton, R. S. {\it et al.} 2017 Nature 541, 58.
\bibitem[\protect\citeauthoryear{Cordes \& Lazio}{2002}]{CL02} Cordes, J. M.
\& Lazio, T. J. W. 2002 arXiv:astro-ph/0207156.
\bibitem[\protect\citeauthoryear{Cordes, Wasserman, Hessels {\it et al.\/}}
{2017}]{CWH17} Cordes, J. M., Wasserman, I., Hessels, J. W. T. {\it et
al.\/} 2017 \apj\ 842, 35.
\bibitem[\protect\citeauthoryear{Dai, Wang, Wu {\it et al.\/}}{2016}]
{DWWH16} Dai, Z. G., Wang, J. S., Wu, X. F. {\it et al.\/} 2016 \apj\ 829,
27.
\bibitem[\protect\citeauthoryear{Farah, Flynn, Bailes {\it et al.\/}}{2018}]
{F18} Farah, W., Flynn, C., Bailes, M. {\it et al.\/} 2018 \mnras\ 478,
1209 arXiv:1803.05697.
\bibitem[\protect\citeauthoryear{Gajjar {\it et al.}}{2018}]{G18} Gajjar,
V., Siemion, A. P. V., MacMahon, D. H. E. {\it et al.\/} 2018 \apj\ 863, 2
arXiv:1804.04101.
\bibitem[\protect\citeauthoryear{Ghisellini \& Locatelli}{2018}]{GL17}
Ghisellini, G. \& Locatelli, N. 2018 \aap\ 613, 61 arXiv:1708.07507.
\bibitem[\protect\citeauthoryear{Gu {\it et al.\/}}{2016}]{GDLMW16} Gu,
W.-M., Dong, Y.-Z., Liu, T. {\it et al.\/} 2016 \apjl\ 823, L28.
\bibitem[\protect\citeauthoryear{Hankins \& Eilek}{2007}]{HE07} Hankins,
T. H. \& Eilek, J. A. 2007 \apj\ 670, 693.
\bibitem[\protect\citeauthoryear{Hardy {\it et al.}}{2017}]{H17} Hardy, L.
K., Dhillon, V. S., Spitler, L. G. {\it et al.} 2017 \mnras\ 472, 2800.
\bibitem[\protect\citeauthoryear{Jackson}{1999}]{J99} Jackson, J. D. 1999
Classical Electrodynamics (3rd ed.) Chichester: Wiley.
\bibitem[\protect\citeauthoryear{Katz}{2014}]{K14} Katz, J. I. 2014 \prd\
89, 103009.
\bibitem[\protect\citeauthoryear{Katz}{2016}]{K16} Katz, J. I. 2016 Mod.
Phys. Lett. A 31, 1630013.
\bibitem[\protect\citeauthoryear{Katz}{2017a}]{K17a} Katz, J. I. 2017a
\mnras\ 467, L96. 
\bibitem[\protect\citeauthoryear{Katz}{2017b}]{K17b} Katz, J. I. 2017b
\mnras\ 469, L39. 
\bibitem[\protect\citeauthoryear{Katz}{2017c}]{K17c} Katz, J. I. 2017c
\mnras\ 471, L92
\bibitem[\protect\citeauthoryear{Keane {\it et al.\/}}{2012}]{Ke12} Keane,
E. F., Stappers, B. W., Kramer, M. {\it et al.\/} 2012 \mnras\ 425, L71.
\bibitem[\protect\citeauthoryear{Kumar, Lu \& Bhattacharya}{2017}]{KLB17}
Kumar, P., Lu, W. \& Bhattacharya, M. 2017 \mnras\ 468, 2726.
\bibitem[\protect\citeauthoryear{Law {\it et al.\/}}{2017}]{L17} Law, C. J.,
Abruzzo, M. W., Bassa, C. G. {\it et al.\/} 2017 \apj\ 850, 76.
\bibitem[\protect\citeauthoryear{Lorimer, {\it et al.\/}}{2007}]{L07}
Lorimer, D. R., Bailes, M., McLaughlin, M. A., Narkevic, D. J. \&
Crawford, F. 2007 \sci\ 318, 777.
\bibitem[\protect\citeauthoryear{Lu \& Kumar}{2018}]{LK17} Lu, W. \& Kumar,
P. 2018 \mnras\ 477, 2470 arXiv:1710.10270.
\bibitem[\protect\citeauthoryear{Marcote {\it et al.}}{2017}]{M17} Marcote,
B., Paragi, Z. Hessels, J. W. T. {\it et al.} 2017 \apjl\ 834, L8.
\bibitem[\protect\citeauthoryear{Masui {\it et al.\/}}{2015}]{M15} Masui,
K., Lin, H.-H., Sievers, J. {\it et al.\/} 2015 Nature 528, 523.
\bibitem[\protect\citeauthoryear{Melrose}{2017}]{Me17} Melrose, D. B. 2017
Rev. Mod. Plasma Phys. 1:5 doi.org/10.1007/s41614-107-0007-0.
\bibitem[\protect\citeauthoryear{Michilli {\it et al.\/}}{2018}]{M18}
Michilli, D., Seymour, A., Hessels, J. W. T. {\it et al.\/} 2018 Nature 553,
182.
\bibitem[\protect\citeauthoryear{Oostrum {\it et al.\/}}{2017}]{O17}
Oostrum, L. C., van Leeuwen, J., Attema, J. {\it et al.\/} 2017 ATel. 10693.
\bibitem[\protect\citeauthoryear{Petroff {\it et al.}}{2015}]{P15} Petroff,
E., Bailes, M., Barr, E. D. {\it et al.\/} \mnras\ 447, 246.
\bibitem[\protect\citeauthoryear{Petroff, Barr, Jameson {\it et al.\/}}
{2016}]{FRBCat} Petroff, E., Barr, E. D., Jameson, A. {\it et al.\/} FRBCAT:
The Fast Radio Burst Catalogue 2016 Pub. Ast. Soc. Aust. 33, id.e045
\url{http://www.frbcat.org} downloaded March 25, 2018.
\bibitem[\protect\citeauthoryear{Petroff {\it et al.}}{2017}]{P17} Petroff,
E., Burke-Spolaor, S., Keane, E. F. {\it et al.\/} 2017 \mnras\ 469, 4465.
\bibitem[\protect\citeauthoryear{Ravi, Shannon \& Jameson}{2015}]{R15} Ravi,
V., Shannon, R. M. \& Jameson, A. 2015 \apjl\ 799, L5.
\bibitem[\protect\citeauthoryear{Ravi {\it et al.\/}}{2016}]{R16} Ravi, V.,
Shannon, R. M., Bailes, M. {\it et al.\/} 2016 \sci\ 354, 1249;
supplementary material in arXiv:1611.05758.
\bibitem[\protect\citeauthoryear{Rickett}{1990}]{R90} Rickett, B. J. 1990
\araa\ 28, 561.
\bibitem[\protect\citeauthoryear{Scholz {\it et al.\/}}{2016}]{Sc16} Scholz,
P., Spitler, L. G., Hessels, J. W. T. {\it et al.\/} 2016 \apj\ 833, 177.
\bibitem[\protect\citeauthoryear{Soglasnov {\it et al.\/}}{2004}]{S04}
Soglasnov, V. A., Popov, M. V., Bartel, N., Cannon, W., Novikov, A. Y.,
Kondratiev, V. I. \& Altunin, V. I. 2004 \apj\ 616, 439.
\bibitem[\protect\citeauthoryear{Spitler {\it et al.\/}}{2014}]{S14}
Spitler, L. G., Cordes, J. M., Hessels, J. W. T. {\it et al.\/} 2014 \apj\
790, 101.
\bibitem[\protect\citeauthoryear{Spitler {\it et al.\/}}{2016}]{Sp16}
Spitler, L.  G., Scholz, P., Hessels, J. W. T. {\it et al.} 2016 Nature 531,
202.
\bibitem[\protect\citeauthoryear{Thornton {\it et al.\/}}{2013}]{T13}
Thornton, D, Stappers, B., Bailes, M. {\it et al.\/} 2013 \sci\ 341, 53.
\bibitem[\protect\citeauthoryear{Wang {\it et al.\/}}{2016}]{WYWDW16} Wang,
J.-S., Yang, Y.-P., Wu, X.-F. {\it et al.\/} 2016 \apjl\ 822, L7.
\bibitem[\protect\citeauthoryear{Yang \& Zhang}{2017}]{YZ17} Yang, Y.-P. \&
Zhang, B. 2017 arXiv:1712.02702.
\end{thebibliography}
\end{document}